\documentclass[twocolumn,showpacs,preprintnumbers,amsmath,amssymb]{revtex4}

\usepackage{graphicx}
\usepackage{dcolumn}
\usepackage{bm}

\begin{document}

\title{Description of Complex Systems in terms of
Self-Organization \\ Processes of Prime Integer Relations}

\author{Victor Korotkikh}
\email{v.korotkikh@cqu.edu.au}

\affiliation{Faculty of Informatics and Communication\\
Central Queensland University\\
Mackay, Queensland 4740, Australia}

\author{Galina Korotkikh}

\email{g.korotkikh@cqu.edu.au}

\affiliation{Faculty of Informatics and Communication\\
Central Queensland University\\
Mackay, Queensland 4740, Australia}

\date{\today}

\begin{abstract}

In the paper we present a description of complex systems in terms
of self-organization processes of prime integer relations. A prime
integer relation is an indivisible element made up of integers as
the basic constituents following a single organizing principle.
The prime integer relations control correlation structures of
complex systems and may describe complex systems in a strong scale
covariant form. It is possible to geometrize the prime integer
relations as two-dimensional patterns and isomorphically express
the self-organization processes through transformations of the
geometric patterns. As a result, prime integer relations can be
measured by corresponding geometric patterns specifying the
dynamics of complex systems. Determined by arithmetic only, the
self-organization processes of prime integer relations can
describe complex systems by information not requiring further
explanations. This gives the possibility to develop an irreducible
theory of complex systems.

\end{abstract}

\pacs{89.75.-k, 89.75.Fb}

\maketitle

\section{Introduction}

Complex systems profoundly change human activities of the day. In
order to understand and control them it becomes increasingly
important to be confident in the theory of complex systems.
Ultimately, this calls for clear explanations why the foundations
of the theory are valid in the first place. The ideal situation
would be to have an irreducible theory of complex systems not
requiring a deeper explanatory base in principle. But the question
arises: where such an irreducible theory may come from, when even
the concept of space-time is questioned \cite{Smolin_1} as a
fundamental entity?

As a possible answer to the question it is suggested that the
concept of integers may take responsibility in the search for an
irreducible theory of complex systems \cite{Korotkikh_1}. The aim
of the paper is to present a description of complex systems in
terms of self-organization processes of prime integer relations.
In particular, it is considered that the prime integer relations
control correlation structures of complex systems and may describe
complex systems in a strong scale covariant form.

A prime integer relation is an indivisible element made up of
integers as the basic constituents following a single organizing
principle. Remarkably, the prime integer relations can be
geometrized as two-dimensional patterns and the self-organization
processes can be isomorphically expressed through transformations
of the geometric patterns. As a result, the self-organization
processes of prime integer relations are characterized
geometrically and quantitatively. In fact, it becomes possible to
measure prime integer relations by corresponding geometric
patterns.

Due to the isomorphism the structure and the dynamics of a complex
system are combined in our description: as a prime integer
relation governs a correlation structure of a complex system, a
corresponding geometric pattern specifies its dynamics.
Applications of the geometrization of prime integer relations are
considered to discuss potential advantages of the proposed
description.

Determined by arithmetic only, the self-organization processes of
prime integer relations can describe complex systems by
information not requiring further explanations. This gives the
possibility to develop an irreducible theory of complex systems.

\section{Invariant Quantities of a Complex System and Underlying
Correlations}

We approach complex systems from a general perspective. In
particular, we describe a complex system by its certain quantities
and are concerned how many of them remain invariant as the system
evolves from one state to another \cite{Korotkikh_1},
\cite{Korotkikh_2}.

Let $I$ be an integer alphabet and
$$
I_{N} = \{x = x_{1}...x_{N}, x_{i} \in I, \ i = 1,...,N \}
$$
be the set of sequences of length $N \geq 2$. We consider a
complex system consisting of $N$ elementary parts with the state
of an elementary part $P_{i}$ specified by a space variable $x_{i}
\in I, \ i = 1,...,N$ and the state of the complex system itself
by a sequence $x = x_{1}...x_{N} \in I_{N}$.

We use piecewise constant functions for a geometric representation
of the sequences. Let $\varepsilon
> 0$ and $\delta > 0$ be length scales of a two-dimensional
lattice. Let $\rho_{m\varepsilon\delta}: x \rightarrow f$ be a
mapping that realizes the geometric representation of a sequence
$x = x_{1}...x_{N} \in I_{N}$ by associating it with a function $f
\in W_{\varepsilon\delta}[t_{m},t_{m+N}]$, denoted $f =
\rho_{m\varepsilon\delta}(x)$, such that
$$f(t_{m}) = x_{1}\delta,
\ f(t) = x_{i}\delta, \ t \in (t_{m+i-1},t_{m+i}], \ i = 1,...,N,
$$
$$
t_{i} = i\varepsilon, \ i = m,...,m + N
$$
and whose integrals $f^{[k]}$ satisfy the condition
$f^{[k]}(t_{m}) = 0, \ k = 1,2,... \ $, where $m$ is an integer.
The sequence $x = x_{1}...x_{N}$ is called a code of the function
$f$ and denoted $c(f)$.

\begin{figure}
\includegraphics[width=.40\textwidth]{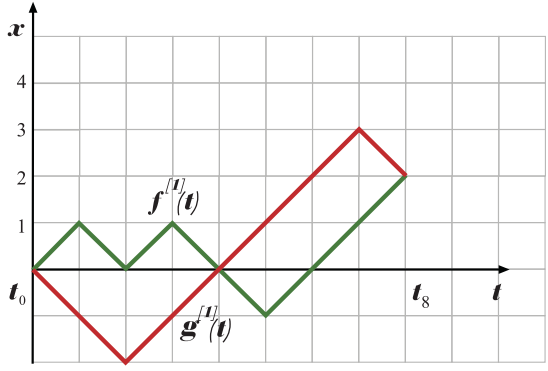}
\caption{\label{fig:one} The figure shows that for states $x =
+1-1+1-1-1+1+1+1$ and $x' = -1-1+1+1+1+1+1-1$ the first integrals
are equal $f^{[1]}(t_{8}) = g^{[1]}(t_{8})$, where $f =
\rho_{m\varepsilon\delta}(x), g = \rho_{m\varepsilon\delta}(x')$
and $m = 0, \varepsilon = 1, \delta = 1$. It turns out that the
second integrals are also equal $f^{[2]}(t_{8}) = g^{[2]}(t_{8})$,
but the third integrals are not $f^{[3]}(t_{8}) \neq
g^{[3]}(t_{8})$. Thus $C(x,x') = 2$.}
\end{figure}

We use the geometric representation to characterize a state $x =
x_{1}...x_{N} \in I_{N}$ of a complex system in terms of the
definite integrals
\begin{equation}
\label{UN1} f^{[k]}(t_{m+N}) = \int_{t_{m}}^{t_{m+N}}
f^{[k-1]}(t)dt, \ k = 1,2,...
\end{equation}
of a function $f^{[0]}= f = \rho_{m\varepsilon\delta}(x) \in
W_{\varepsilon\delta}[t_{m},t_{m+N}]$ \cite{Korotkikh_1}. The
definite integrals give us certain quantities of the complex
system.

Remarkably, the integer code series \cite{Korotkikh_3} expresses
the quantities $(\ref{UN1})$ of a complex system in terms of the
state $x = x_{1}...x_{N} \in I_{N}$ explicitly. In particular, it
describes the definite integral
\begin{equation}
\label{UN2} f^{[k]}(t_{m+N}) = \sum_{i = 0}^{k-1}a_{kmi}
((m+N)^{i}x_{1} + ... + (m+1)^{i}x_{N})\varepsilon^{k}\delta
\end{equation}
of a function $f \in W_{\varepsilon\delta}([t_{m},t_{m+N}])$ by
using the code $c(f) = x_{1}...x_{N}$ of the function $f$, powers
$$
(m+N)^{i},...,(m+1)^{i}, \ i = 0,...,k-1
$$
of integers $(m+N),...,(m+1)$ and combinatorial coefficients
$$a_{kmi} =
((-1)^{k-i-1}(m+1)^{k-i}+(-1)^{k-i}m^{k-i})/(k-i)!i!,
$$ where $k
\geq 1$ and $i = 0,...,k-1$ \cite{Korotkikh_3}.

We are concerned how many quantities $(\ref{UN1})$ remain
invariant as a complex system moves from one state $x =
x_{1}...x_{N} \in I_{N}$ at a time $\tau$ to another state $x' =
x_{1}'...x_{N}' \in I_{N}$ at a time $\tau'$ \cite{Korotkikh_1},
\cite{Korotkikh_2}
\begin{equation}
\label{UN3} f^{[k]}(t_{m+N}) = g^{[k]}(t_{m+N}), \ \ \ k = 1,...,
C(x,x'),
\end{equation}
\begin{equation}
\label{UN4} f^{[C(x,x')+1]}(t_{m+N}) \neq
g^{[C(x,x')+1]}(t_{m+N}),
\end{equation}
where $f = \rho_{m\varepsilon\delta}(x), g =
\rho_{m\varepsilon\delta}(x')$ (Figure 1).

We consider the conservation of the quantities $(\ref{UN3})$ in
view of $(\ref{UN4})$ as a consequence of the correlations between
the parts of the complex system. Although the complex system moves
from one state $x$ to another $x'$, yet the changes of the parts
are correlated to preserve $C(x,x')$ of the quantities. The
complex system can be characterized by the rate of change of the
correlations with respect to time.

\section{The Conserving Correlations as Specific Linear Equations}

By using $(\ref{UN2})$ it is proved that $C(x,x') \geq 1$ of the
quantities $(\ref{UN1})$ of a complex system remain invariant, as
it moves from one state $x = x_{1}...x_{N} \in I_{N}$ to another
$x' = x_{1}'...x_{N}' \in I_{N}$, iff the space-time changes
$\Delta x_{i} = x_{i}' - x_{i}, i = 1,...,N$ of the elementary
parts satisfy a system of $C(x,x')$ linear equations
\cite{Korotkikh_1}
$$
(m + N)^{0}\Delta x_{1} + ... + (m + 1)^{0}\Delta x_{N} = 0
$$
$$
. \qquad \qquad . \qquad  \qquad .  \qquad \qquad . \qquad  \qquad
. \qquad \qquad .
$$
\begin{equation}
\label{UN5} (m + N)^{C(x,x')-1}\Delta x_{1} + ... + (m +
1)^{C(x,x')-1}\Delta x_{N} = 0
\end{equation}
and an inequality
\begin{equation}
\label{UN6} (m + N)^{C(x,x')}\Delta x_{1} + ... + (m +
1)^{C(x,x')}\Delta x_{N} \neq 0.
\end{equation}

The coefficients of the system of linear equations $(\ref{UN5})$
become the entries of the matrix
$$
\left(\begin{array}{ccccc}
(m + N)^{0} & (m + N-1)^{0} & ... & (m + 1)^{0}\\
(m + N)^{1} & (m + N-1)^{1} & ... & (m + 1)^{1}\\
. & ... &  & . \\
(m + N)^{N-1} & (m + N-1)^{N-1} & ... & (m + 1)^{N-1}\\
\end{array} \right)
$$
with the Vandermonde determinant, when the number of the equations
is $N$. This fact is important in order to prove that the number
$C(x,x')$ of the conserved quantities of a complex system
satisfies the condition $C(x,x') < N$ \cite{Korotkikh_1}.

The system of equations $(\ref{UN5})$ may bring interesting
associations. For example:

1. It is suggested that the system of equations $(\ref{UN5})$ may
be connected with a smooth projective curve over a finite field
and the number $C(x,x')$ of the conserved quantities with the
genus of the curve \cite{Korotkikh_1}.

2. The system of equations $(\ref{UN5})$ can be written for $s =
0,-1,...,-C(x,x')+1$ and $m = 0$ as
$$
\sum_{n = 1}^{N} \frac{\Delta x_{N-n+1}}{n^{s}} = 0
$$
to have a naive resemblance with the Dirichlet zeta function
$$
L(s,\chi) = \sum_{n = 1}^{\infty}\frac{\chi_{n}}{n^{s}},
$$
where $\chi_{n}$ are some coefficients and $s$ defined for proper
complex numbers. This points to a possible link of the equations
$(\ref{UN5})$ with the zeroes of the Dirichlet zeta function
$$
L(s,\chi) = \sum_{n = 1}^{\infty}\frac{\chi_{n}}{n^{s}} = 0
$$
and the zeroes of the Riemann zeta function
$$
\zeta(s) = \sum_{n = 1}^{\infty}\frac{1}{n^{s}} = 0
$$
in particular.

\section{Self-Organization Processes of Prime Integer Relations and
Correlation Structures of Complex Systems}

The analysis of the equations $(\ref{UN5})$ and inequality
$(\ref{UN6})$ reveals that a complex system is actually described
in terms of hierarchical structures of prime integer relations.
Such a hierarchical structure can be interpreted as a result of a
self-organization process of prime integer relations
\cite{Korotkikh_1}. Namely, the self-organization process,
starting with integers as the elementary building blocks and
following a single principle, makes up the prime integer relations
of one level of the hierarchical structure from the prime integer
relations of the lower level.

\begin{figure}
\includegraphics[width=.50\textwidth]{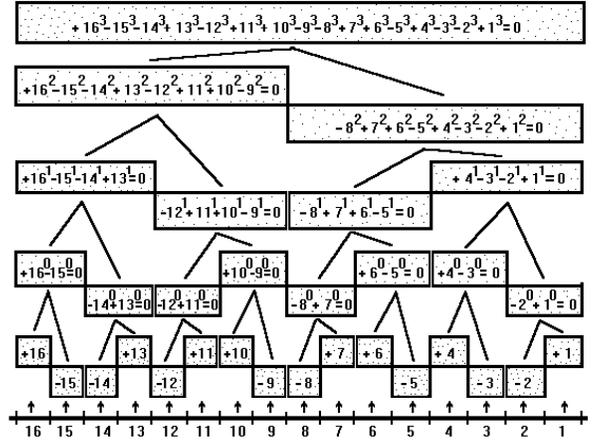}
\caption{\label{fig:two} One of the hierarchical structures
associated with the integer relations $(\ref{UN7})$ and inequality
$(\ref{UN8})$. The structure can be interpreted as a result of a
self-organization process of prime integer relations. The process
starts with integers $1, 4, 6, 7, 10, 11, 13, 16$ in the positive
state and integers $2, 3, 5, 8, 9, 12, 14, 15$ in the negative
state at the zero level to make up the prime integer relations at
the first level. Guided by a single principle, the
self-organization process forms the prime integer relations from
the first to the fourth level. But the process can not reach the
fifth level, because, according to arithmetic, the left side of
$(\ref{UN8})$ does not equal zero.}
\end{figure}

We illustrate the results by considering two states of a complex
system
$$
x = -1+1+1-1+1-1-1+1+1-1-1+1-1+1+1-1,
$$
$$
x' = +1-1-1+1-1+1+1-1-1+1+1-1+1-1-1+1
$$
specified by the Prouhet-Thue-Morse (PTM) sequences of length $N =
16$. In this case since $C(x,x') = 4$, the equations $(\ref{UN5})$
can be written as four integer relations
$$
+16^{0} - 15^{0} - 14^{0} + 13^{0} - 12^{0} + 11^{0} + 10^{0} -
9^{0}
$$
$$
- 8^{0} + 7^{0} + 6^{0} - 5^{0} + 4^{0} - 3^{0} - 2^{0} + 1^{0} =
0,
$$
$$
+16^{1} - 15^{1} - 14^{1} + 13^{1} - 12^{1} + 11^{1} + 10^{1} -
9^{1}
$$
$$
- 8^{1} + 7^{1} + 6^{1} - 5^{1} + 4^{1} - 3^{1} - 2^{1} + 1^{1} =
0,
$$
$$
+16^{2} - 15^{2} - 14^{2} + 13^{2} - 12^{2} + 11^{2} + 10^{2} -
9^{2}
$$
$$
- 8^{2} + 7^{2} + 6^{2} - 5^{2} + 4^{2} - 3^{2} - 2^{2} + 1^{2} =
0,
$$
$$
+16^{3} - 15^{3} - 14^{3} + 13^{3} - 12^{3} + 11^{3} + 10^{3} -
9^{3}
$$
\begin{equation}
\label{UN7} - 8^{3} + 7^{3} + 6^{3} - 5^{3} + 4^{3} - 3^{3} -
2^{3} + 1^{3} = 0
\end{equation}
and the inequality $(\ref{UN6})$ takes the form
$$
+16^{4} - 15^{4} - 14^{4} + 13^{4} - 12^{4} + 11^{4} + 10^{4} -
9^{4}
$$
\begin{equation}
\label{UN8} -8^{4} + 7^{4} + 6^{4} - 5^{4} + 4^{4} - 3^{4} - 2^{4}
+ 1^{4} \neq 0,
\end{equation}
where $m = 0$ and a common factor $2$ originated from the
space-time variables $\Delta x_{i}, i = 1,...,16$ is not shown.
This leaves the dynamics of the elementary parts represented by
the signs only, but allows us to focus on the integer relations
and inequality, and identify prime integer relations.

It is worth to note that calculations in $(\ref{UN7})$ and
$(\ref{UN8})$ and their results are completely determined by
arithmetic.

There is a number of hierarchical structures of prime integer
relations associated with the system of integer relations
$(\ref{UN7})$ and inequality $(\ref{UN8})$. One of the
hierarchical structures is shown in Figure 2. In the structure the
relationships between the elements of neighboring levels can be
interpreted as a consequence of a self-organization process of
prime integer relations. The process starts with integers
$1,...,16$ in certain states, i.e., positive or negative, and
proceeds level by level following the same organizing principle:
\smallskip
\smallskip

{\it on each level the powers of the integers in the prime integer
relations are increased by $1$, so that through emerging
arithmetic interdependencies the prime integer relations could
self-organize as the components to form prime integer relations of
the higher level.}
\smallskip

It is important to note that the formation of prime integer
relations is more than their simple sum.

Let us mention again that there are many hierarchical structures
of prime integer relations associated with $(\ref{UN7})$ and
$(\ref{UN8})$. To indicate where their variety comes from we give
an alternative hierarchical structure of prime integer relations.
The first level of this structure includes the prime integer
relations
$$
+16^{0} - 14^{0} = 0, \ \ \ -15^{0} + 13^{0} = 0, \ \ \ -12^{0} +
10^{0} = 0,
$$
$$
+11^{0} - 9^{0} = 0, \ \ \ -8^{0} + 6^{0} = 0, \ \ \  +7^{0} -
5^{0} = 0,
$$
$$
+4^{0} - 2^{0} = 0, \ \ \ -3^{0} + 1^{0} = 0.
$$

The second level - the prime integer relations
$$
(+16^{1} - 14^{1}) + (-15^{1} + 13^{1}) = 0,
$$
$$
(-12^{1} + 10^{1}) + (+11^{1} - 9^{1}) = 0,
$$
$$
(-8^{1} + 6^{1}) + (+7^{1} - 5^{1}) = 0, \ \ \ (+4^{1} - 2^{1}) +
(-3^{1} + 1^{1}) = 0.
$$

The third level - the prime integer relations
$$
((+16^{2} - 14^{2}) + (-15^{2} + 13^{2}))
$$
$$
+ ((-12^{2} + 10^{2}) + (+11^{2} - 9^{2})) = 0,
$$
$$
((-8^{2} + 6^{2}) + (+7^{2} - 5^{2})) + ((+4^{2} - 2^{2}) +
(-3^{2} + 1^{2})) = 0.
$$

The fourth level - the prime integer relation
$$
(((+16^{3} - 14^{3}) + (-15^{3} + 13^{3}) + ((- 12^{3} + 10^{3}) +
(+11^{3} - 9^{3})))
$$
$$
+ (((-8^{3} + 6^{3}) + (+7^{3} - 5^{3})) + ((+4^{3} - 2^{3}) + (-
3^{3} + 1^{3}))) = 0.
$$

Let us explain the notion of prime integer relation. A prime
integer relation of the first level is made up of integers from
the zero level. An integer comes to the first level in positive or
negative state (Figure 2). Following the organizing principle
prime integer relations of a level make up a prime integer
relation of the higher level as an indivisible element. Namely, if
even one of the prime integer relations is not involved, then
according to the organizing principle the rest of the prime
integer relations can not form an integer relation.

By our definition an integer relation
$$
+7^{2} - 6^{2} - 5^{2} + 3^{2} + 2^{2} - 1^{2} = 0
$$
is a prime integer relation. However, an integer relation
$$
+7^{0} - 6^{0} - 5^{0} + 3^{0} = 0
$$
is not prime, because it consists of two prime integer relations
$+7^{0} - 6^{0} = 0, \  -5^{0} + 3^{0} = 0$. For simplicity prime
integer relations such as
$$
+14^{0} - 7^{0} = 0, \ \ \ +2 \cdot 14^{0} - 2 \cdot 7^{0} = 0
$$
are not distinguished. Multiple $2$ means that we have two
integers $14$ in the positive state and two integers $7$ in the
negative state.

The space-time dynamics $\Delta x_{i}$ of an elementary part
$P_{i}, i = 1,...,N$ has the following interpretation. The
absolute value of $\Delta x_{i}$ is the number of integers $(m + N
- i +1)$ starting each of the self-organization processes
associated with the system of equations $(\ref{UN5})$ and
inequality $(\ref{UN6})$, while $sign(\Delta x_{i})$ determines
the state, i.e., positive or negative, of the integers, provided
that $\Delta x_{i} \neq 0, i = 1,...,N$.

The correlation structures underlying the conservation of the
quantities $(\ref{UN3})$ in view of $(\ref{UN4})$ are defined by
the hierarchical structures of prime integer relations associated
with the system of equations $(\ref{UN5})$ and inequality
$(\ref{UN6})$ \cite{Korotkikh_1}, \cite{Korotkikh_4},
\cite{Korotkikh_5}.

We illustrate the result by using the system of integer relations
$(\ref{UN7})$, where we return to the symbolic form of
$(\ref{UN5})$ to change our perspective from the self-organization
processes of prime integer relations to the dynamics of the
complex system in space-time. Starting with the prime integer
relation $+16^{0} - 15^{0} = 0$ (Figure 2), which actually stands
as
$$
+16^{0}(+2) + 15^{0}(-2) =
$$
\begin{equation}
\label{UN9}
+16^{0}\Delta x_{1} + 15^{0}\Delta x_{2} = 0,
\end{equation}
we can see that space-time changes $\Delta x_{1}$ and $\Delta
x_{2}$ of the elementary parts $P_{1}$ and $P_{2}$ are correlated
as
\begin{equation}
\label{UN10}
+16^{0}\Delta x_{1} = - 15^{0}\Delta x_{2}
\end{equation}
and the elementary parts $P_{1}$ and $P_{2}$ thus make a composite
part $(P_{1} \leftrightarrow P_{2})$.

The prime integer relation $(\ref{UN9})$ does not contain
information about a physical signal that may realize the
correlation between the parts $P_{1}$ and $P_{2}$. But, if the
dynamics $\Delta x_{2}$ of the elementary part $P_{2}$ is
specified, then, according to $(\ref{UN10})$, the dynamics $\Delta
x_{1}$ of the elementary part $P_{1}$ is instantaneously
determined and vice versa. We may also specify that the
correlation is nonlocal, because the prime integer relation
$(\ref{UN9})$ does not have any reference to the distance between
the parts $P_{1}$ and $P_{2}$.

\begin{figure}
\includegraphics[width=.48\textwidth]{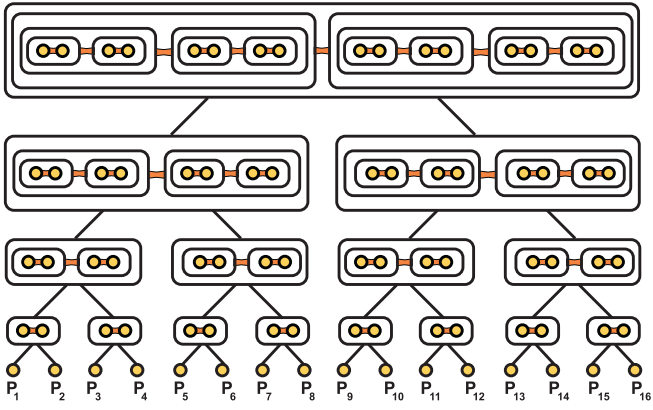}
\caption{\label{fig:three} A correlation structure of the complex
system is defined by the hierarchical structure of prime integer
relations (Figure 2). A horizontal link denotes that the parts are
correlated through a prime integer relation. As arithmetic behind
the prime integer relations makes them sensitive to a minor
change, so does the correlation structure. If the complex system
deviates from the dynamic behavior even slightly, then some of the
correlation links disappear and the complex system decays.}
\end{figure}

Similarly, the prime integer relation $-14^{0} + 13^{0} = 0$ leads
to
$$
+14^{0}\Delta x_{3} + 13^{0}\Delta x_{4} = 0,
$$
which specifies the correlation
$$
14^{0}\Delta x_{3} = - 13^{0}\Delta x_{4}
$$
between the elementary parts $P_{3}$ and $P_{4}$ and describes a
composite part $(P_{3} \leftrightarrow P_{4})$.

In its turn the prime integer relation
$$
+16^{1} - 15^{1}- 14^{1} + 13^{1} = 0,
$$
made up of the prime integer relations $+16^{0} - 15^{0} = 0$ and
$-14^{0} + 13^{0} = 0$, corresponds to
$$
(16^{1} \Delta x_{1} + 15^{1} \Delta x_{2}) + (14^{1}\Delta x_{3}
+ 13^{1} \Delta x_{4}) = 0,
$$
which shows that the composite parts $(P_{1} \leftrightarrow
P_{2})$ and $(P_{3} \leftrightarrow P_{4})$ are correlated as
$$
(16^{1} \Delta x_{1} + 15^{1}\Delta x_{2}) = - (14^{1}\Delta x_{3}
+ 13^{1}\Delta x_{4})
$$ and form a larger composite part
$$
((P_{1} \leftrightarrow P_{2}) \leftrightarrow (P_{3}
\leftrightarrow P_{4})).
$$

Continuing the consideration we can associate the
self-organization process of prime integer relations with the
formation of a correlation structure of the complex system
(Figures 2 and 3). The formation process starts with the
elementary parts $P_{1},...,P_{16}$ and combine them into
composite parts to make up then larger composite parts and so on
until the whole correlation structure is built.

A complex system can be described by self-organization processes
of prime integer relations in a distinctive way. Information about
a complex system can be given by prime integer relations, which
are true statements not requiring further explanations. The prime
integer relations are organized as hierarchical structures and
there is no need for deeper principles to explain why the
hierarchical structures exist the way they do and not otherwise.

For example, there is no need to explain a hierarchical structure,
where prime integer relations
$$
+7^{0}-6^{0}=0, \ \ \ -5^{0}+3^{0}=0, \ \ \ +2^{0}-1^{0}= 0
$$
of level $1$ form a prime integer relation
\begin{equation}
\label{UN11}
+7^{1}-6^{1}-5^{1}+3^{1}+2^{1}-1^{1}= 0
\end{equation}
of level $2$. The prime integer relation $(\ref{UN11})$ alone
makes up a prime integer relation
\begin{equation}
\label{UN12}
+7^{2}-6^{2}-5^{2}+3^{2}+2^{2}-1^{2} = 0
\end{equation}
of level $3$. However, the prime integer relation $(\ref{UN12})$
on its own can not progress to level $4$, because
$$
+7^{3}-6^{3}-5^{3}+3^{3}+2^{3}-1^{3}\neq 0.
$$

By using the self-organization processes of prime integer
relations a concept of complexity is introduced \cite{Korotkikh_1}
and its applications may be found in \cite{Korotkikh_6}.

\section{Geometrization of the Self-Organization Processes of
Prime Integer Relations and its Applications}

Prime integer relations as abstract entities are not natives of
mental pictures associated with formations of physical objects. At
the same time in our description of complex systems prime integer
relations behave like objects that under the control of arithmetic
can transform into each other. This view would be more relevant
with prime integer relations as geometric objects suitable for
measurement to obtain information about complex systems.

Remarkably, by using the integer code series \cite{Korotkikh_3},
the prime integer relations can be geometrized as two-dimensional
patterns and the self-organization processes can be isomorphically
expressed through transformations of the geometric patterns
\cite{Korotkikh_1}. As a result, the self-organization processes
of prime integer relations are characterized geometrically and
quantitatively. In fact, it becomes possible to measure prime
integer relations by corresponding geometric patterns.

Due to the isomorphism the structure and the dynamics of a complex
system are combined in our description: as a prime integer
relation governs a correlation structure of a complex system, a
corresponding geometric pattern specifies its dynamics. The
geometrization of the prime integer relations puts forward
arithmetic to irreducibly explain complex systems through
quantitative means.

To illustrate the results we consider a self-organization process
of prime integer relations that can progress through the
hierarchical levels \cite{Korotkikh_1}. The process is connected
with the PTM sequence and can be specified in terms of critical
point \cite{Kadanoff_1}, \cite{Wilson_1} features
\cite{Korotkikh_1}. This resonates with the fact that the PTM
sequence gives a symbolic description of chaos resulting from the
period-doubling \cite{Feigenbaum_1} in a complex system
\cite{Allouche_1}.

\begin{figure}
\includegraphics[width=.50\textwidth]{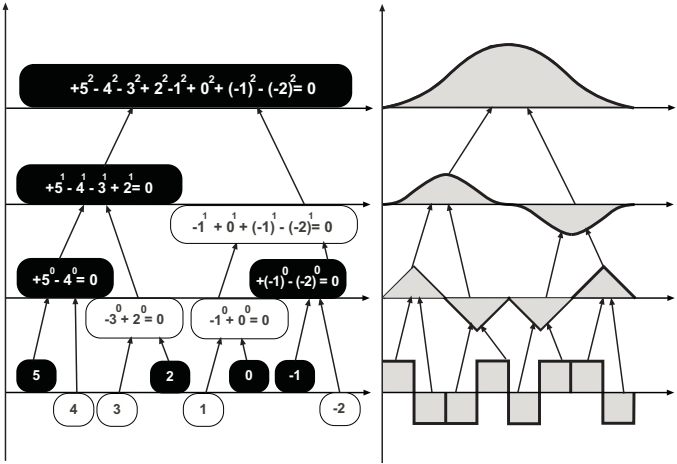}
\caption{\label{fig:four} A hierarchical structure of prime
integer relations and an isomorphic hierarchical structure of
geometric patterns. A prime integer relation can be positive
(shown in black) and negative (shown in white). Under the
integration of the function, the geometric patterns of one level
form the geometric patterns of the higher level, so we can observe
the geometric patterns length scale by length scale. A prime
integer relation can be measured by the area of a corresponding
geometric pattern or the length of its boundary curve. The
boundary curve of a geometric pattern specifies the dynamics of an
associated complex system.}
\end{figure}

The left side of Figure 4 shows one of the hierarchical structures
of prime integer relations, when
$$
x = 0 \ \ 0 \ \ 0 \ \ 0 \ \ 0 \ \ 0 \ \ 0 \ \ 0,
$$
$$
x' = +1 -1 -1 +1 -1 +1 +1 -1,
$$
$m = -3$ and a complex system consists of $N = 8$ elementary parts
$P_{i}, i = 1,...,8$. The sequence $x'$ is the initial segment of
length $8$ of the PTM sequence starting with $+1$.

The right side of Figure 4 presents an isomorphic hierarchical
structure of geometric patterns as a result of the geometrization
of the self-organization process of prime integer relations. The
geometrization allows us to visualize the process and investigate
it, as well as the complex system, geometrically and
quantitatively. In particular, the self-organization process can
be analyzed by measuring prime integer relations through their
geometric patterns. For example, a prime integer relation can be
measured by the area of a corresponding geometric pattern or the
length of its boundary curve. These two characteristics of a prime
integer relation are quantities of an associated complex system.
We consider the quantities in applications to discuss potential
advantages of the proposed description.

The first application shows that complex systems may be described
by the prime integer relations in a strong scale covariant form.
Namely, although a PTM geometric pattern at level ${\cal N}
> 1$ of length scale $2^{{\cal N}}\varepsilon$ is bounded by
an intricate curve, nevertheless the PTM geometric pattern has a
concise and universal description working for all levels. In
particular, the area $S$ of a PTM geometric pattern at level
${\cal N} \geq 1$ can be expressed as if it were a triangle
\begin{equation}
\label{UN13} S = \frac{LH}{2},
\end{equation}
where $L$ and $H$ are the length and the height of the PTM
geometric pattern (Figure 4). Consequently, the law of PTM pattern
area is the same for all length scales and in the simplest
possible form. In other words, the description of $S$ is {\it
strongly scale covariant}, i.e.,
\smallskip

{\it under the scale transformations the equation of $S$ is
preserved in the simplest possible form $(\ref{UN13})$.}
\smallskip

A PTM geometric pattern is a result of the formation, but for the
description of its area $S$ there is no need to know what happens
at the lower levels. All information can be obtained by measuring
the length $L$ and the height $H$ of the PTM geometric pattern at
the level of consideration. Nevertheless, the history of the
formation of a PTM geometric pattern is encoded by the boundary
curve.

Because of $(\ref{UN13})$ the PTM geometric patterns have a scale
invariant property that divides the hierarchical levels into
groups of {\it three} successive levels. Namely, the lengths and
the heights of PTM geometric patterns at levels ${\cal N} = 1,2,3$
and ${\cal N} = 4,5,6$ are given in terms of $\varepsilon$ and
$\delta$ as
$$
(2\varepsilon, \varepsilon\delta), \ (4\varepsilon,
\varepsilon^{2}\delta), \ (8\varepsilon, 2\varepsilon^{3}\delta)
$$ and
$$
(16\varepsilon, 8\varepsilon^{4}\delta), \ (32\varepsilon,
64\varepsilon^{5}\delta), \
(64\varepsilon,1024\varepsilon^{6}\delta).
$$

By using the renomalization group transformation
$$
\varepsilon' = 2^{3}\varepsilon, \ \delta' =
\varepsilon^{3}\delta,
$$
the lengths and the heights of PTM geometric patterns at levels
${\cal N} = 4,5,6$ can be expressed in terms of $\varepsilon'$ and
$\delta'$ in the same way
$$
(2\varepsilon', \varepsilon'\delta'), \ (4\varepsilon',
\varepsilon'^{2}\delta'), \ (8\varepsilon',
2\varepsilon'^{3}\delta')
$$
as the lengths and the heights of PTM geometric patterns at levels
${\cal N} = 1,2,3$ are given in terms of $\varepsilon$ and
$\delta$. The situation repeats for levels ${\cal N} = 7,8,9$ and
so on.
\smallskip

The second application is connected with a long-standing problem
to explain why constants of nature, such as the fine-structure
constant $\alpha$ measured to be equal to $1/137.03599976$ and
written as
$$
\alpha = 0.00729735... \ \ ,
$$
have the values they do and not even slightly different. Although
the logic of digits in the fine-structure constant $\alpha$ has
not been established yet, it is known that $\alpha$ is fragile. If
the fine-structure constant $\alpha$ varied even a bit, then
complex physical systems would not be able to exist
\cite{Barrow_1}.

A prime integer relation is also an intricate entity, because
arithmetic behind makes it sensitive to a minor change of the
elements. However, when we read a prime integer relation, unlike a
constant of nature, we can understand and accept it as a true
statement not requiring further explanations. Moreover, a prime
integer relation can describe a correlation structure of the
complex system. In this capacity it encodes the parts of the
correlation structure, the relationships between them, i.e., how
the parts are connected, and the strengths of the relationships,
i.e., how the dynamics of some parts of a relationship determine
the dynamics of the other parts (Figures 2 and 3). In our
description a minor change breaking a prime integer relation also
leads to a collapse of a corresponding complex system, because
some of the relationships between the parts of the correlation
structure disappear.

The analogy with the sensitivity of the fine-structure constant
$\alpha$ could be made stronger, if a prime integer relation would
be found sensitive to a single number. The geometrization of a
prime integer relation provides a number with the required
property. In particular, a prime integer relation can be
isomorphically expressed as a two-dimensional geometric pattern,
which is completely determined by the boundary curve (Figure 4).
The length of the curve, i.e., a number, encodes the geometric
pattern and thus the prime integer relation. With even a minor
change to the number the prime integer relation as well as a
corresponding complex system cease to exist. Indeed, if the number
i.e., the length, changes even slightly, then the boundary curve
in its turn changes the geometric pattern, which leads the prime
integer relation and the complex system to decay.

We consider the example above with $m = 0, \varepsilon = 1$ and
$\delta = 1$ (Figure 4) to show how such numbers can be obtained.
By using corresponding geometric patterns at levels 1, 2 and 3, we
define the numbers for prime integer relations.

For the prime integer relation $+8^{0} - 7^{0} = 0$ we have
$$
\vartheta_{1} = 2 \int_{0}^{1}\sqrt{1 + (\frac{df^{[1]}}{dt})^{2}}
dt =
$$
$$
2 \int_{0}^{1}\sqrt{1 + 1}dt =  2\sqrt{2} = 2 \times
1.414223562... \ ,
$$
where $f = \rho_{011}(x'), x' = +1-1-1+1-1+1+1-1$.

For the prime integer relation
$$
8^{1} - 7^{1} -6^{1} + 5^{1}= 0
$$
we obtain
$$
\vartheta_{2} = 4 \int_{0}^{1}\sqrt{1 + (\frac{df^{[2]}}{dt})^{2}}
dt =
$$
$$
4 \int_{0}^{1} \sqrt{1 + t^{2}}dt = 4 \times 1.14779... \ .
$$

For the prime integer relation
$$
8^{2} - 7^{2} - 6^{2} + 5^{2} - 4^{2} + 3^{2} + 2^{2} - 1^{2}= 0
$$
we get
$$
\vartheta_{3} = 4 \int_{0}^{2}\sqrt{1 + (\frac{df^{[3]}}{dt})^{2}}
dt =
$$
$$
4 (\int_{0}^{1} \sqrt{1 + \frac{t^{4}}{4}}dt + \int_{1}^{2}
\sqrt{1 + (\frac{-t^{2}}{2} + 2t -1)^{2}}dt) =
$$
$$
4 \times (1.0242... + 1.30702...) = 4 \times 2.33122... \ .
$$

The numerical results in the second and third cases are computed
by using {\it Mathematica}.

We can write the prime integer relations and their corresponding
numbers as
$$ +8^{0} - 7^{0} = 0 \ \ \ \Longrightarrow 2 \times
1.414223562...
$$
$$
+8^{1} - 7^{1} -6^{1} + 5^{1}= 0 \ \ \ \Longrightarrow 4 \times
1.14779...
$$
$$
+8^{2} - 7^{2} - 6^{2} + 5^{2} - 4^{2} + 3^{2} + 2^{2} - 1^{2}= 0
\ \ \ \Longrightarrow 4 \times 2.33122... \ .
$$

On the side of prime integer relations we have confidence in the
arithmetic statements, as we can check them. Moreover, we know how
the prime integer relations are built, can observe symmetry in
their corresponding geometric patterns (Figure 4) and associate
them with correlation structures of complex systems.

But on the other side it is not clear what logic digits of the
numbers may follow. The situation would become intriguing, once
the numbers turned to be found as constants by physical
experiments.

When a constant of nature is measured, the information from
physical devices comes for processing in the numerical form.
Presenting numerical information through prime integer relations
may give us a tool to understand experimental results and make our
decisions relying on irreducible arguments. For instance, if some
physical experiments identify a constant and as a number it
corresponds in our description to a prime integer relation
$$
+7^{2}-6^{2}-5^{2}+3^{2}+2^{2}-1^{2} = 0,
$$
then it may be considered that the experiments through the
constant actually reveal a true statement we can understand and
agree with.
\smallskip

Therefore, we propose to explore the idea:
\smallskip

{\it constants of nature may be numerical expressions of prime
integer relations or their metrics}.
\smallskip

If it were the case, then constants of nature would be understood
without further explanations.

\section{Conclusions}

We have shown that complex systems can be described in terms of
self-organization processes of prime integer relations. The
processes have the integers as the basic building blocks and
controlled by arithmetic only make up the prime integer relations
from one level to the higher level. In the description of complex
systems prime integer relations demonstrate remarkable properties:

- following a single principle may self-organize at one level to
form a prime integer relation at the higher level. A prime integer
relation may participate in various self-organization processes
and, as a result, be a component of different prime integer
relations at the higher level;

- through the self-organization processes the prime integer
relations are interconnected and inseparable in one hierarchical
network. Not even a minor change can be made to any element of the
network;

- control correlation structures of complex systems and may
describe complex systems in a strong scale covariant form. Prime
integer relations specify nonlocal and instantaneous correlations;

- provide a complexity order. The self-organization processes
start with different integers and, as a result, progress to
different levels thus producing the complexity order. It seems
like a self-organization process of prime integer relations aims
to progress as higher as possible in the direction of the order;

- can be geometrized and may be measured in renormalizable numbers
by corresponding geometric patterns. The self-organization
processes of prime integer relations can be isomorphically
expressed through transformations of the geometric patterns. The
geometrization of the prime integer relations puts forward
arithmetic to irreducibly explain complex systems through
quantitative means;

- determined by arithmetic only, the self-organization processes
of prime integer relations can describe complex systems by
information not requiring further reductions. This property could
be particularly useful as irreducible arguments may be needed to
explain the fundamental laws of complex physical systems
\cite{Weinberg_1}.

Finally, we have presented self-organization processes of prime
integer relations as a new way to describe complex systems. The
processes can characterize complex systems by information not
requiring further explanations. This gives the possibility to
develop an irreducible theory of complex systems.
\smallskip

This work was supported by CQU Research Advancement Awards Scheme
grants no. IN9022.

\bibliography{apssamp}

\begin{thebibliography}{00}

\bibitem{Smolin_1} L. Smolin, {\it Three Roads to Quantum Gravity}
(Basic Books, 2001).

\bibitem{Korotkikh_1} V. Korotkikh, {\it A Mathematical Structure for
Emergent Computation} (Kluwer, Dordrecht, 1999).

\bibitem{Korotkikh_2} V. Korotkikh, {\it Integers: Irreducible Guides
in the Search for a Unified Theory}, Braz. J. Phys, Special Issue
on Decoherence, Information, Complexity and Entropy, {\bf 35}(2B),
509 (2005).

\bibitem{Korotkikh_3} V. Korotkikh, {\it Integer Code Series with Some
Applications in Dynamical Systems and Complexity} (Computing
Centre of the Russian Academy of Sciences, Moscow, 1993).

\bibitem{Korotkikh_4} G. Korotkikh and V. Korotkikh, in {\it Optimization and
Industry: New Frontiers}, ed. by P. Pardalos and V. Korotkikh
(Kluwer, Dordrecht, 2003).

\bibitem{Korotkikh_5} V. Korotkikh, in {\it Fuzzy Partial Differential
Equations and Relational Equations}, ed. by M. Nikravesh, L. Zadeh
and V. Korotkikh (Springer, Berlin, 2004).

\bibitem{Korotkikh_6} V. Korotkikh, G. Korotkikh and D. Bond,
{\it On Optimality Condition of Complex Systems: Computational
Evidence}, arXiv:cs/0504092, April, 2005.

\bibitem{Kadanoff_1} L. Kadanoff, W. Gotze, D. Hamblen, R. Hecht,
E.A.S. Lewis, V.V. Palciaukas, M. Rayl and J. Swift, Rev. Mod.
Phys. {\bf 39}(2), 395 (1967).

\bibitem{Wilson_1} K. G. Wilson and J. Kogut, Phys. Rep. C{\bf 12}, 75
(1974).

\bibitem{Feigenbaum_1} M. Feigenbaum, Los Alamos Sci. {\bf 1},
4 (1980).

\bibitem{Allouche_1} J. Allouche and M. Cosnard, in {\it Dynamical
Systems and Cellular Automata} (Academic Press, 1985).

\bibitem{Barrow_1} J.D. Barrow, {\it The Constants of Nature: From
Alpha to Omega} (Jonathan Cape, London, 2002) and (Pantheon, New
York, 2002); J.D. Barrow and J.K. Webb, Sci. Amer. {\bf 292}, 6,
33 (2005).

\bibitem{Weinberg_1} S. Weinberg, {\it Dreams of a Final Theory}
(Vintage Books, New York, 1992).


\end{thebibliography}

\end{document}